\documentclass[12pt,a4]{article}
\usepackage{psfig,epsfig,graphicx}
\usepackage{amsmath}
\usepackage{amssymb}
\usepackage{cite}

\begin{document}

\title{Shear Viscosity of Hot QED at Finite Density from Kubo Formula}
\author{Liu Hui \thanks{liuhui@iopp.ccnu.edu.cn} \ \ Hou Defu \thanks{hdf@iopp.ccnu.edu.cn}
 \ \ Li Jiarong \thanks{ljr@iopp.ccnu.edu.cn} \\[0.5cm] {\it\small Institute of Particle Physics, Central China Normal University,}\\ {\it
\small Wuhan(430079), P.R.China}}
\date{}
\maketitle
\begin{abstract}
Within the framework of finite temperature field theory  this
paper discusses the shear viscosity of hot QED plasma through Kubo
formula at one-loop skeleton diagram level with a finite chemical
potential. The effective widths(damping rates) are introduced to
regulate the pinch singularities. The finite chemical potential,
which enhances the contributions to the shear viscosity from the
electrons while suppresses those from the photons, finally gives a
positive contribution compared to the pure temperature
environment. The result agrees with that from the kinetics theory
qualitatively.
\end{abstract}

\section{Introduction}
Transport properties of relativistic plasma are of great interest
both experimentally and theoretically. Taking the signal of quark
gluon plasm(QGP) for example, the so-called 'strong-coupled
matter' formed in the relativistic heavy ion collider(RHIC) was
well described by an ideal hydrodynamics model in the region of
$p_T<2GeV$ to fit the data of the elliptic flow
$v_2$\cite{star,phenix}. But it is pointed out that the
over-prediction in high-$p_t$ region of such model is due to the
neglect of dissipation and viscosity\cite{Molnar,Teaney}. Thereby
the transport coefficients need to be considered. In addition, the
non-ideal fluid properties of viscosity and dissipation may
influence some other important equations and quantities such as
the equation of state(EoS), the formation(thermalization) time and
the sound attenuation length
etc.\cite{Teaney,Blaizot,Moore,shuryak}.

From the theoretical points of view, two formalisms are usually
developed to calculate the transport coefficients. One is kinetic
theory which starts from expanding the distribution function near
the local equilibrium when dealing with the Boltzmann equation.
The transport coefficients, especially the shear viscosity, were
discussed by many authors in this
formalism\cite{Hosoya,Gavin,Danielewicz,Oertzen,Baym,Arnold1,Arnold2}.
The other way of obtaining the transport coefficients is to
implement the Kubo formulae within the framework of the finite
temperature field theory. The Kubo formulae, which work in both
weak coupling and strong coupling regime, have been applied for
computing the transport coefficients  of strong-coupled theory in
lattice Monte Carlo simulations\cite{Karsch,Gupta,Nakamura}, and
for analytical calculation in scalar theory\cite{jeon1,wang,wang2}
and gauge theory\cite{Thoma,defu,Aarts,Aarts2,Basagoiti,defu2},
early with relaxation time approximation and later with ladder
diagram resummation and large $N_f$ expansion technique. However
these two approaches are not irrelevant. Their inner relations are
attractive theoretically. Some literatures have shown that the
Kubo formulae with ladder resummation are equivalent to the
kinetics theory in scalar field and gauge
field\cite{jeon1,Jeon,Carrington,Basagoiti}. In principle, one
should resum various diagrams with ladder vertices and effective
propagators when evaluating the correlation functions in Kubo
formulae, which leads to a set of complicated integral equations.
In this paper, we develop another approach, distinguished from the
ladder resummation and the naive one-loop calculation which gives
an inconvincible result, trying to demonstrate that even at
one-loop skeleton diagram level the Kubo formula is still
consistent qualitatively with kinetics theory if only the proper
effective photon and fermion width are chosen as the infrared
regulators in which the main physical characters are involved.

In heavy-ion physics, the chemical potential in the central fire
ball is not zero thought small compared to the extremely high
temperature environment. And most of the existing publications
concentrated only on the high temperature case but neglected the
small chemical potential in the heavy ion collision. To involve
the chemical potential effect is our main motivation. It makes
sense to introduce a small chemical potential $\mu(\mu\ll T)$ to
study that to what extent it can affect the properties of the
plasma. The transport coefficients of QGP at finite density has
been discussed in Ref. \cite{defu} with the relaxation-time
approximation, and early estimated by Danielewicz and Gyulassy to
scale the hydrodynamical equations\cite{Danielewicz}. We estimated
the shear viscosity of hot QED plasma at finite temperature and
chemical potential, referring to the results in the kinetics
framework, since the infrared pinch singularities will appear and
long distant interaction should be considered in the transport
processes.

The whole context will be arranged as follows: In section 2, we
review the Kubo formula relevant to shear viscosity in the finite
temperature theory. The contributions from the photons and the
electrons in the QED plasma are studied respectively in section 3
and 4, in which one can see the transport damping rates play
important roles in regulating the pinch singularities. Conclusions
and discussions are presented in section 5.

Here are some notations in this paper: (a) the capital letter
stands for the four-momentum, $K=(k_0, {\mathbf{k}})$ with
$k=|{\mathbf{k}}|$. (b) $\int \frac{d^4 K}{(2\pi)^4}$ is
abbreviated to $\int dK$. (c) When we mention 'hard', it means the
momentum is much larger than $T$, or at least at the same order;
'soft' means much smaller than $eT$ or at least at the same order.

\section{Shear Viscosity and Two-point Retarded Green Function}

Generally speaking, the longitudinal expansion domains in heavy
ion collision, therefore the shear viscosity is more important
than the bulk viscosity which is proved much smaller than the
former in lattice calculation\cite{Nakamura}. In this paper, we
ignore the bulk viscosity and just take the shear viscous one into
account.

In a near equilibrium system with linear response theory, Kubo
formula tells us the shear viscosity\cite{Carrington} is
determined by
\begin{equation}
\eta=\frac{1}{10}\int d^3x'\int^0_{-\infty}dt'G_R(0;x',t'),
\end{equation}
where $G_R(0;x',t')$ is the two-point retarded Green function
defined as:
\begin{equation}
G_R(x,t;x',t')=-i\theta(t-t')<[\pi_{ij}(x,t),\pi_{ji}(x',t')]>,\label{rgf}
\end{equation}
and $\pi_{ij}$ is the spacial part of the dissipative
energy-momentum tensor
\begin{equation}
\pi_{ij}=(\delta_{ik}\delta_{jl}-\frac{1}{3}\delta_{ij}\delta_{kl})T^{kl}=-\eta\left(\partial_{i}u_{j}-\partial_{j}u_{i}+\frac{2}{3}\delta_{ij}\partial_k
u^k\right),
\end{equation}
where $T^{kl}$ is the spacial energy momentum. The coefficients
$\eta$ is the shear viscosity. The four-velocity $u^{\mu}(x)$ in
the local rest frame is $(1,0,0,0)$.

Transforming  Eq.(\ref{rgf}) into momentum space, one obtains
\begin{equation}
\eta=-\frac{i}{10}\frac{d}{dq_0}[\ \lim\limits_{q\rightarrow
0}G_R(Q)\ ]_{q_0=0},\label{v}
\end{equation}
with $G_R(Q)=\int dX' e^{-iQ\cdot X'}G_R(0;t',x')$ and
$X'=(t',x')$.

Through the Kubo formula we know that the shear viscosity is
determined by the two-point retarded Green function. Now we start
to evaluate it in an explicit dynamics. In the relativistic heavy
ion collision, people are interested in the extremely high
temperature environment, therefore one can use hard thermal
loop(HTL) approximation theoretically. We here choose QED as a
sample model and calculate the contributions both from the bosons
and from the fermions.

The QED Lagrangian reads:
\begin{equation}
\mathcal{L}_{QED}=\bar\Psi(i\gamma^\mu\cdot\partial_\mu-m)\Psi-\frac{1}{4}F^{\mu\nu}F_{\mu\nu}-e\bar\Psi\gamma^\mu\Psi
A_\mu,
\end{equation}
where $F_{\mu\nu}=\partial_\mu A_\nu-\partial_\nu A_\mu$ and
$\Psi, \bar\Psi, A_\mu$ are the electron, anti-electron and photon
fields respectively.

With this Lagrangian, one can read the viscous stress tensors  in
the momentum space,
\begin{equation}
\pi^e_{ij}(k)=i \bar\Psi(\gamma_i
k_j-\frac{1}{3}\delta_{ij}{\vec\gamma}\cdot{\mathbf{k}})\Psi,\label{vf}
\end{equation}
for electrons and
\begin{equation}
\pi^\gamma_{ij}(k)=-(k_i k_j-\frac{1}{3}\delta_{ij}k^2)A^2\label{vb}
\end{equation}
for gauge boson with the Lorentz gauge condition.

Combining Eqs.(\ref{rgf}), (\ref{v}), (\ref{vf}) and (\ref{vb}),
with the standard definitions of propagator in quantum field
theory and transport vertex defined behind, one can figure out the
one-loop skeleton diagrams of the retarded Green function
presented in Fig.1.
\begin{figure}
 \begin{center}
   \resizebox{8cm}{!}{\includegraphics{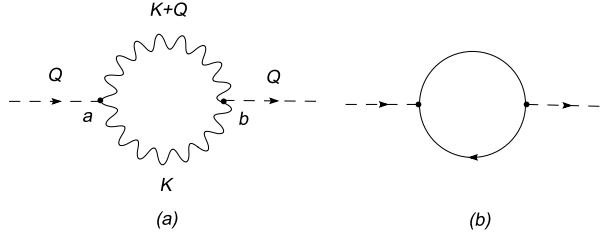}}
 \end{center}
   \caption{Two point retarded Green function of boson(a) and fermion(b). The dashed line represents the
   spacial stress tensor operator $\pi_{ij}$, the solid line for fermions and
the wiggle line for bosons.}
\end{figure}
The main difference between the propagators of the normal field
and the thermal field rest on the fact that the latter is a
$2\times2$ matrix due to the doubled Hilbert
space\cite{Carrington2,Kapusta2,LeBellac}.

In a high-temperature environment, the loop momenta are hard
therefore the mass of fermions can be omitted and bare thermal
propagators are sufficient. Yet in the straightforward calculation
pinch singularities occur when the external four-momentum
approaches to zero, which is required by Eq.(\ref{v}). To avoid
such divergences, we must keep the imaginary part of the
propagator, namely, the effective damping rate(widths), as
infrared regulators, which can be obtained analytically at
leading-log order.

\section{Boson Contributions}
The diagram of the boson two-point Green function is demonstrated
in Fig.1(a), with which one can expressed it as
\begin{equation}
G^{ph}_{ab}(Q)=i\int dK D^{\mu\nu}_{ab}(K)I_{ij}^\gamma \tau_a
          D_{ba,\mu\nu}(K+Q)I^\gamma_{ji}\tau_b,
\end{equation}
where $D_{ab}^{\mu\nu}(Q)$ is the thermal gauge boson propagator,
and a, b=1, 2 are the doubled Hilbert space indices;
$\tau_a,\tau_b=\pm 1$ represent the type-1 and type-2 vertices
respectively. The notation  $I^\gamma_{ij}=-k_i
k_j+\frac{1}{3}\delta_{ij}k^2$ is the transport vertex of the
photon viscosity. Summing over the Lorentz indices and using the
reverse relation of
 Keldysh representation\cite{Carrington2,Keldysh} we obtain:
\begin{eqnarray}
G^{ph}_R(K)&=&G^{ph}_{11}(Q)-G^{ph}_{12}(Q)\label{bs}\\
      &=&\frac{32i}{3}\int dK
      k^4[n_B(k_0)-n_B(k_0+q_0)]D_A(K)D_R(K+Q),\nonumber
\end{eqnarray}
in which $n_B(k_0)=(e^{\beta |k_0|}-1)^{-1}$ is the Bose-Einstein
distribution.  $D_{R,A}(K)$ is the retarded(advanced) gauge boson
propagator in the Feynman gauge with
$D^{\mu\nu}_{R,A}(K)=g^{\mu\nu}D_{R,A}(K)$ and
\begin{equation}
D_{R,A}(K)=\frac{1}{K^2\pm i \varepsilon sgn(k_0)}.
\end{equation}
In the last line of Eq.(\ref{bs}), the fluctuation-dissipation
theorem has been employed:
\begin{equation}
G_S(P)=[1+2n_B(p_0)]sgn(p_0)[D_R(P)-D_A(P)].
\end{equation}

 Here we encounter the pinch problem of $D_A(K)D_R(K+Q)$ in
Eq.(\ref{bs}) when the external momentum $Q$  approaches to zero
which is  required by the definition of viscosity. The physical
reason responsible for this divergence is the naive adoption of
bare propagators. Therefore the so-called HTL resummed propagator
developed by Pisarski {\it et al.} needs to be considered to
involve the medium effect, namely, the scheme of eliminating the
pinch is to endow the finite particle life-time or width
consistently. Mathematically, we insert a small $\gamma_{ph}$ into
the imaginary part of the propagator which physically means the
damping rate. The real part of self-energy(thermal mass term),
which is in $gT$ order, can be neglected compared with the hard
loop momentum. We drop out $Q$ in the propagators directly since
Eq.(\ref{v}) requires $q_0=0$ after
 derivation on $q_0$, and thus the thermal distribution in Eq.(\ref{bs}) has  the form
\begin{equation}
n_B(k_0)-n_B(k_0+q_0)\approx q_0\beta n_B(k_0)[1+n_B(k_0)]+\cdots
\end{equation}
which preserves the first order in $q_0$. This procedure can be
written manifestly as
\begin{eqnarray}
D_A(K)D_R(K+Q)&&\xrightarrow{Q\rightarrow
0}\frac{1}{(k_0+i\gamma_{ph})^2-k^2}\cdot\frac{1}{(k_0-i\gamma_{ph})^2-k^2}\\
&&\longrightarrow
\frac{\pi}{4k^2\gamma_{ph}}[\delta(k_0-k)+\delta(k_0+k))].\label{delta}
\end{eqnarray}
Notice that the photon damping rate is of order $e^4 T$
 , much smaller than the momentum $K\gtrsim T$, which implies the width of the
 photon can be substituted by a delta function approximately in
  expression (\ref{delta}). Integrating off the delta function and the
azimuth angles in the spherical coordinates, the retarded Green
function becomes
\begin{equation}
G^{ph}_R(Q)=\frac{8i\beta q_0}{3\pi^2}\int dk k^4
n_B(k)[1+n_B(k)]\frac{1}{\gamma_{ph}}.\label{bosongreen}
\end{equation}

Now we are in the position to calculate the photon damping rate
consistently at a finite temperature and density. There is a fact
one should notice that the longitudinal contribution is suppressed
compared to the transverse one for hard momenta \cite{LeBellac}.
To the leading order evaluation, one should only consider the
latter. It was calculated by Thoma with vanishing chemical
potential \cite{Thoma3}, and we will generalize it to the finite
density case.

The photon damping in the plasma owes to two processes, Compton
scattering and pair production(Fig.2). The two diagrams in the
first line of fig.2 stand for Compton scattering and the last line
is for pair production processes.

\begin{figure}
 \begin{center}
   \resizebox{10cm}{!}{\includegraphics{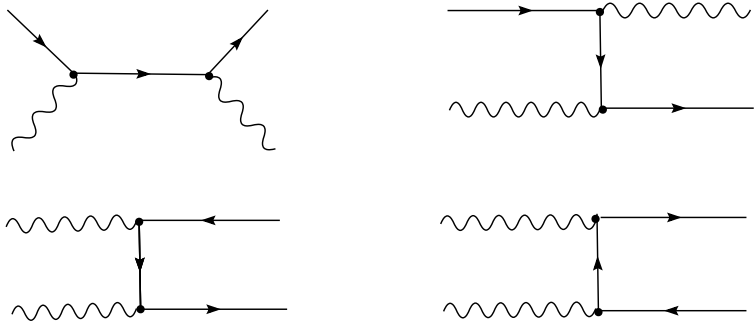}}
 \end{center}
   \caption{The physical processes contribute to the photon damping rate. The first line is for the Compton scattering and the last line for
   the pair production.}
\end{figure}
The Braaten-Pisarski method\cite{Bratten2} has been used to
calculate the photon damping rate in the QED plasma, where the
total effect is divided into two parts by a separating scale. The
soft contribution is obtained by cutting the photon self-energy
diagram with the HTL resummed fermion propagator(Fig.3) and the
hard contribution can be evaluated most conveniently from the
matrix elements. Adding up the two contributions one can cancel
the separating scale.

\subsection{Soft Contributions}
The soft contribution of a real hard photon damping rate follows
from the the imaginary part of transverse self-energy in Fig.3,
where the blob denotes the full fermion propagator.

\begin{figure}
 \begin{center}
   \resizebox{10cm}{!}{\includegraphics{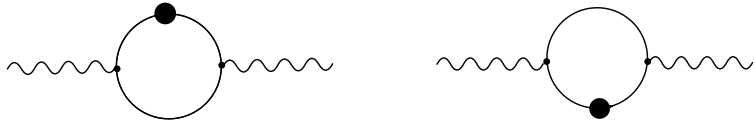}}
 \end{center}
   \caption{Hard photon self-energy with the full propagators denoted by a blob.}
\end{figure}
The damping rate of a photon is defined as
\begin{equation}
\gamma=-\frac{1}{2p}Im\Pi_T(p_0=p,p),
\end{equation}
in the case of non-overdamping. The self-energy is given by
\begin{equation}
\Pi_{\mu\nu}=2ie^2\int dK Tr[S^*(K)\gamma_\mu
S(P-K)\gamma_\nu]\label{photonselfenergy}
\end{equation}
where $S^*(K)$ is the effective fermion propagator, the factor 2
accounts for both diagrams in Fig.3. We take $Q\equiv P-K$ and
adopt the helicity representation of the fermion propagator,
\begin{eqnarray}
S^*(K)&=&\frac{1}{D_+(K)}\frac{\gamma_0-\hat{\mathbf{k}}\cdot \vec
\gamma}{2}+\frac{1}{D_-(K)}\frac{\gamma_0+\hat{\mathbf{k}}\cdot
\vec
\gamma}{2}\\
S(Q)&=&\frac{1}{d_+(Q)}\frac{\gamma_0-\hat{\mathbf{q}}\cdot \vec
\gamma}{2}+\frac{1}{d_-(Q)}\frac{\gamma_0+\hat{\mathbf{q}}\cdot
\vec \gamma}{2},
\end{eqnarray}
where $\hat{\mathbf{k}} =\mathbf{k}/k$ and
\begin{eqnarray}
D_{\pm}(K)&=&-k_0\pm k+\frac{m^2}{2k}[(1\mp\frac{k_0}{k})\ln\frac{k_0+k}{k_0-k}\pm 2 ] \\
d_{\pm}(Q)&=&-q_0\pm q,
\end{eqnarray}
with $z=k_0/k$. Here the full fermion propagator is the result
from HTL resummation. Projecting the self-energy on the transverse
direction, one finds
\begin{equation}
\Pi_T(p_0,p)=\frac{1}{2}(\delta_{ij}-\frac{p_ip_j}{p^2})\Pi_{ij}(p_0,p).
\end{equation}
Evaluating the trace over $\gamma$ matrices, one can transform the
transverse part of photon self-energy\cite{Kapusta} into
\begin{equation}
\Pi_T(P)=2ie^2\int dK \left[\frac{1}{D_+(K)}\left
(\frac{1-V}{d_+(Q)}+\frac{1+V}{d_-(Q)}\right
)+\frac{1}{D_-(K)}\left
(\frac{1+V}{d_+(Q)}+\frac{1-V}{d_-(Q)}\right)\right ]
\end{equation}
where $V=(\mathbf{\hat p}\cdot\mathbf{\hat k})(\mathbf{\hat p}
\cdot \mathbf{\hat q})$.

To sum the frequencies over the fourth momenta with a chemical
potential, we first prove an auxilary formula which is
 useful to sum over the Matusubara frequencies in a typical
one loop diagram at finite chemical potential. The zero chemical
potential case was first developed by Braaten, Pisarski and
Yuan\cite{Braaten}.

The typical one loop diagram  one encounters in the temperature
field theory may take the form of
\begin{equation}
O(i\omega)=T\sum\limits_n
\triangle(i\omega_n+\mu)\triangle(i\omega
-i\omega_n-\mu)\label{oneloop}
\end{equation}
where $i\omega$ is the external line momentum and $i\omega_n=2\pi
(n+1)/\beta$ are the fermion Matusubara frequencies. Employing the
spectral representation\cite{LeBellac},
\begin{equation}
\triangle(i\omega_n+\mu)=-\int dk_0
\frac{\rho(k_0)}{i\omega_n+\mu-k_0},
\end{equation}
 one recasts Eq.(\ref{oneloop}) into
\begin{equation}
O(i\omega)=\int dk_0 \int dq_0 \rho(k_0)\rho'(q_0)T\sum\limits_n
\frac{1}{i\omega_n-k_0+\mu} \cdot
\frac{1}{i\omega-i\omega_n-k_0+\mu}.
\end{equation}
Summing over the Matusubara frequencies $n$, one obtains
\begin{equation}
O(i\omega)=-\int dk_0 \int dq_0 \rho(k_0)\rho'(q_0)
\frac{1-f_+(k_0)-f_-(q_0)}{i\omega-k_0-q_0}
\end{equation}
with $f_{\pm}(k_0)=[e^{\beta(k_0\mp\mu)}+1]^{-1}$.

Let $i\omega$ be analytically continued to $p_0$ which are not the
Matusubara frequencies and the imaginary part of $O(p_0)$ is
\begin{eqnarray}
\mbox{Im} O(p_0)&=&\frac{1}{2i}\mbox{Disc}
O(p_0)=\frac{1}{2i}[O(p_0+i\epsilon)-O(p_0-i\epsilon)] \\
       &=&\pi(1-e^{\beta p_0})\int dk_0 \int dq_0
       f_+(k_0)f_-(q_0)\rho(k_0)\rho'(q_0) \delta(p_0-k_0-q_0)).\nonumber
\end{eqnarray}
When $T\gg K\gg\mu$, $f_+(k_0)f_-(q_0)=f_+(k_0)f_-(p_0-k_0)\sim
e^{-\beta p_0}/2$,
\begin{equation}
\mbox{Im}O(p_0)=\frac{\pi}{2}(1-e^{\beta p_0})e^{-\beta p_0}\int
dk_0 \int dq_0 \rho(k_0)\rho(q_0)\delta(p_0-k_0-q_0).\label{imsum}
\end{equation}

Using the generalized Eq.(\ref{imsum}) and adopting the similar
steps in Ref. \cite{Kapusta}, we finally obtain
\begin{equation}
\gamma_{soft}=\frac{\pi}{4p}\alpha^2\left(T^2+\frac{\mu^2}{\pi^2}\right)\ln\frac{\Lambda^2}{\pi\alpha(T^2+\frac{\mu^2}{\pi^2})}\label{soft}
\end{equation}
where $\alpha=e^2/4\pi$ is the fine structure constant, and
$\Lambda$ is the separation scale, namely, the upper limit of
three-momentum integration.

\subsection{Hard Contributions} The hard contribution, when the
exchanged fermion carries momentum much larger than the separating
scale $\Lambda$, can be evaluated through the matrix elements. The
damping rate relevant to the Compton scattering process is
\begin{eqnarray}
\gamma^{comp}_{hard}=&&\frac{1}{4p}\int \frac{d^3k}{(2\pi)^3
2k}n_F(k)\frac{d^3p}{(2\pi)^3 2p}[1+n_B(p')]\int
\frac{d^3k'}{(2\pi)^3
2k'}[1-n_F(k')]\nonumber\\
&&\times (2\pi)^4\delta^4(P+K-P'-K')4<|\mathcal M|^2>_{comp},
\end{eqnarray}
where  $<|M|>_{comp}$ is the scattering amplitude and
$n_F(k)=(e^{\beta |k_0|}+1)^{-1}$ is the Fermi-Dirac distribution.
 Similarly, the damping rate for pair creation process is
\begin{eqnarray}
\gamma^{pair}_{hard}=&&\frac{1}{4p}\int \frac{d^3k}{(2\pi)^3
2k}n_B(k)\frac{d^3p}{(2\pi)^3 2p}[1-n_F(p')]\int
\frac{d^3k'}{(2\pi)^3
2k'}[1-n_F(k')]\nonumber\\
&&\times (2\pi)^4\delta^4(P+K-P'-K')2<|\mathcal M|^2>_{pair}.
\end{eqnarray}

Using Mandelstam variables $s=(P+K)^2, t=(P-P')^2$ and $u=-s-t$,
one finds the matrix elements are given by\cite{Halzen}
\begin{eqnarray}
<|\mathcal M|^2>_{comp}&=&-2e^4\left
(\frac{u}{s}+\frac{s}{u}\right ),\\
<|\mathcal M|^2>_{pair}&=&2e^4\left (\frac{u}{t}+\frac{t}{u}\right
).
\end{eqnarray}

In the hard external lines assumption, that is $P'\gg T$ and
$K'\gg T$, the phase space for the outside region is unfavorable,
thus the distribution functions can be simplified as $1\pm
n_{F,B}\approx 1$. This considerable simplification may bring us
much convenience when using the Lorentz invariant phase space
factor\cite{Halzen}
\begin{eqnarray}
dL&=&(2\pi)^4\delta^4(P+K-P'-K')\frac{d^3p'}{(2\pi)^3
2p'}\frac{d^3k'}{(2\pi)^3 2k'}\nonumber \\
&=&\frac{dt}{8\pi s}.
\end{eqnarray}

In our approximation, the density effect, namely the chemical
potential, modifies only the Compton scattering process due to the
initial fermion distribution, while those in pair production are
kept unchanged as in the pure temperature environment. And the
result is already given by Ref. \cite{Thoma3}
\begin{equation}
\gamma^{pair}_{hard}=\frac{\pi}{6}\frac{\alpha^2
T^2}{p}(\ln\frac{4pT}{\Lambda^2}-2.1472).\label{pair}
\end{equation}
In the Compton process, considering the initial fermion
distribution function with a small chemical potential, we can
obtain the corresponding damping rate
\begin{equation}
\gamma^{comp}_{hard}=\frac{\pi}{12p}\alpha^2(T^2+\frac{3\mu^2}{\pi^2})(\ln\frac{4pT}{\Lambda^2}-0.0772)-\frac{\alpha^2T^2}{2\pi
p}A(\mu)\label{comp}
\end{equation}
where
\begin{equation}
A(\mu)=S_2(e^{-\beta\mu})+S_2(e^{\beta\mu}),
\end{equation}
\begin{equation}
S_n(z)=\sum\limits_{k=1}\limits^{\infty}\frac{z^k}{k^n}\ln k, \ \
\ \ \ \ n=0,1,2 \cdots.
\end{equation}
Adding up Eqs.(\ref{pair}) and (\ref{comp}), one obtains the hard
contribution,
\begin{equation}
\gamma_{hard}=\frac{\pi}{4p}\alpha^2\left
[(T^2+\frac{\mu^2}{\pi^2})\ln\frac{4pT}{\Lambda^2}-(1.4572+\frac{2A(\mu)}{\pi^2})T^2-0.0078\mu^2
\right ].\label{hard}
\end{equation}

When the soft and the hard contributions are combined together,
the separation scale $\Lambda$ is to be cancelled thus yielding
\begin{eqnarray}
\gamma&=&\gamma_{soft}+\gamma_{hard}\nonumber \\
      &=&\frac{\pi}{4p}\alpha^2(T^2+\frac{\mu^2}{\pi^2})F(T,\mu)\label{photondamping}
\end{eqnarray}
with
\begin{equation}
F(T,\mu)=\ln\frac{\langle p\rangle
T}{\pi\alpha\left(T^2+\frac{\mu^2}{\pi^2}\right)}-\frac{1}{T^2+\frac{\mu^2}{\pi^2}}\left
[\left(1.4572+\frac{2A(\mu)}{\pi^2}\right)T^2+0.0078\mu^2 \right
].
\end{equation}
In the above expression of $F(T,\mu)$ we replaced the momentum in
the logarithm by its average value $\langle p\rangle$ to simplify
the future integration, which is defined as
\begin{equation}
\langle p\rangle =\frac{\int d^3k \ k \ n_B(k)}{\int d^3k\
n_B(k)}=2.701T.
\end{equation}

Inserting the Eq.(\ref{photondamping}) into (\ref{bosongreen}),
completing the three momentum integration and with Eq.(\ref{v}),
we finally obtain:
\begin{equation}
\eta_{ph}=\frac{1.072T^5}{\alpha^2(T^2+\frac{\mu^2}{\pi^2})F(T,\mu)}\approx
84.64\frac{T^3}{e^4\ln e^{-1}}(1-0.101\frac{\mu^2}{T^2}).
\end{equation}
where the last approximation is obtained with expansion in terms
of $\mu^2/T^2$ and keeping solely the leading-log accuracy.

\section{Fermion Contribution} In this section we are going to
calculate the shear viscosity contributed from the electrons. With
Eq. (\ref{vf}) and Fig. 1(b), one can write
\begin{equation}
G^e_{ab}(Q)=i\int dK Tr[S_{ab}(K\!\!\!\!\! \slash\ )I_{ij}^e
\tau_b S_{ab}(K\!\!\!\!\! \slash+Q\!\!\!\! \slash\
)I_{ji}^e\tau_a],
\end{equation}
where $I^e_{ij}=\gamma_i
k_j-\frac{1}{3}\delta_{ij}\vec\gamma\cdot\mathbf{k}$ and
$S_{ab}(K\!\!\!\! \slash)$ is the element of $2\times2$ fermion
propagator matrix. Summed over the thermal indices, the retarded
Green function then becomes:
 \begin{eqnarray}\label{fermiongreen}
 G^e_R(Q)&=&G_{11}(Q)-G_{12}(Q)\\
       &=&i\int dK Tr[S_{11}(K\!\!\!\!\! \slash\ )I_{ij}^e S_{11}(K\!\!\!\!\! \slash+Q\!\!\!\! \slash\ )I_{ji}^e
       -S_{12}(K\!\!\!\!\! \slash\ )I_{ij}^e S_{21}(K\!\!\!\!\! \slash+Q\!\!\!\! \slash\ )I_{ji}^e]\nonumber .
 \end{eqnarray}
Using the inverse relation of Keldysh representation and the
fluctuation-dissipation theorem, the last line of Eq.
(\ref{fermiongreen}) recasts into
\begin{equation}
i\int dK Tr[K\!\!\!\!\! \slash\ I_{ij}^e(K\!\!\!\!\!
\slash+Q\!\!\!\! \slash\ )I_{ji}^e)][\tilde n_F(k_0)-\tilde
n_F(k_0+q_0)] \tilde{S_R}(K)\tilde{S_A}(K+Q),\label{dr}
\end{equation}
where $\tilde n_F(k_0)=[e^{\beta(|k_0|\pm \mu)}+1]^{-1}$ and
$S(K\!\!\!\!\! \slash\ )=K\!\!\!\!\! \slash \ \tilde{S}(K)$.

We notice that
\begin{equation}
Tr[K\!\!\!\!\! \slash\ I_{ij}^e(K\!\!\!\!\! \slash+Q\!\!\!\!
\slash\ )I_{ji}^e)]=\frac{8}{3}k^4,
\end{equation}
and
\begin{equation}
\tilde{S_R}(K)\tilde{S_A}(K+Q)\xrightarrow{Q\rightarrow
0}\frac{1}{K^2+i\varepsilon sgn(k_0)}\cdot\frac{1}{
K^2-i\varepsilon sgn(k_0)}. \label{fs}
\end{equation}
Similarly, considering the fermion damping in the plasma and using
the electron damping rate $\gamma_e$ to improve the infrared
behavior, one obtains
\begin{equation}
\tilde{S_R}(K)\tilde{S_A}(K+Q)\xrightarrow{Q\rightarrow
0}\frac{1}{(k_0+i\gamma_e)^2-k^2}\cdot\frac{1}{(k_0-i\gamma_e)^2-k^2}.\label{fs2}
\end{equation}

\begin{figure}
 \begin{center}
   \resizebox{6cm}{!}{\includegraphics{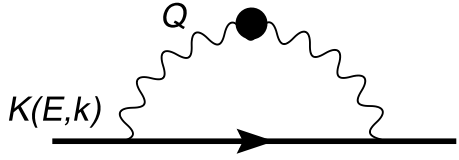}}
 \end{center}
   \caption{Hard fermion damping in the QED plasma. The blob on the photon line stands by
   a resumed propagator.}
\end{figure}

The electron damping rate, defined by the imaginary part of the
electron self-energy on mass shell with full photon
propagator(Fig.4), is just one half of the interaction rate
$\Gamma$. According to the cutting rules\cite{Weldon,Braaten3},
the interaction rate can be described by the photon longitudinal
and transverse spectrum functions $\rho_l$ and $\rho_t$ defined in
the HTL approximation for space-like frequencies as
\begin{equation}
\Gamma(E)=\frac{e^2}{2\pi v }\int^\infty_0 dq\ q
\int^{vq}_{-vq}\frac{d\omega}{\omega}\left[\rho_l(\omega,q)+(1-\frac{\omega^2}{q^2})\rho_t(\omega,q)\right]
\end{equation}
with $v\equiv k/E$ and
\begin{eqnarray}
\rho_l(\omega,q)&=&\frac{3m_\gamma^2\omega}{2q}\left[\left(q^2+3m_\gamma^2-\frac{3m_\gamma^2\omega}{2q}\log\frac{q+\omega}{q-\omega}\right)^2+\left(\frac{3\pi
m_\gamma^2\omega}{2q}\right)^2\right]^{-1}\\
\rho_t(\omega,q)&=&\frac{3m_\gamma^2\omega(q^2-\omega^2)}{4q^3}\left\{\left[q^2-\omega^2+\frac{3m_\gamma^2\omega}{2q^2}\left(1+\frac{q^2-\omega^2}{2\omega
q}
\log\frac{q+\omega}{q-\omega}\right)\right]^2\right.\nonumber\\
&& \ \ \ \ \ \ \ \ \ \ \ \ \ \ \ \ \ \ \ \ \
\left.\left(\frac{3m_\gamma^2\omega(q^2-\omega^2)}{4q^3}\right)^2\right\}^{-1}
\end{eqnarray}
where
$m_\gamma^2=\frac{1}{9}e^2T^2(1+\frac{3}{\pi^2}\frac{\mu^2}{T^2})$
is the photon thermal mass. Performing integrals of $dq$ and
$d\omega$ with quasi-static approximation, i.e. $v\ll 1$, one
 finds that the longitudinal part is screened by the
thermal mass while the transverse part suffers a logarithmic
infrared divergency due to the absence of magnetic screening.
Introducing an infrared cutoff $\epsilon$ by hand, one can count
the power dependency of the coupling constant
\begin{equation}\label{fermiondamping1}
\Gamma(E)\sim e^2T(\ln \epsilon^{-1}+constant),
\end{equation}
which makes the viscosity $\eta\sim e^{-2}$, namely, the
reciprocal of quadratic coupling constant, but not quartic as
kinetic theory suggusts\cite{Hosoya,Arnold1,Liu}. This
disagreement hints us to discover the defects in our previous
calculation.

Returning back to Fig.(1), one would find that only involving
effective propagator is not enough, which means the rungs between
the two propagators have been omitted. To include such diagrams we
must sum up all these ladder diagrams. While that is a rather
complex procedure and a set of integral equations is obtained as a
result\cite{jeon1,Jeon,Carrington,Basagoiti}. Here we do not
employ this scheme, instead we replace the fermion damping rate,
namely the infrared regulator, with a more physical transport
damping rate, which has been applied even in the discussion of
classic system and Abelian or non-Abelian plasma transport
coefficients\cite{Danielewicz,Thoma,defu,Aarts2,xuzhe}.

For a Rutherford-like scattering in classic plasma, namely, the
elastic scattering with the same type and momentum of incoming and
outgoing particles, small angle scattering domains in the
dynamical cross section. Whereas, as we know, the large angle
scattering, which is caused by a series of accumulated multiple
small angle scattering in a long distance, is the most effective
to the shear viscous process. Thus an additional factor of
$1-\cos\theta$ is multiplied in front of the ordinary cross
section\cite{lifshitz}, where $\theta$ is the scattering angle. In
this procedure the so-called transport cross section is defined.
Furthermore, when discussing the transport process in QED or QCD
plasma with the Boltzmann equation in kinetics theory, one will
find that the collision term is the convolution integral of
scattering amplitudes and their correspondent statistical weights,
where if the interaction vertex connects the two particles of the
same type, the fluctuation part of the statistic distribution
functions will contribute an extra small $q^2$\cite{Arnold1},
\begin{equation}\label{kinetics factor q}
[\chi_{ij}(\mathbf{p+q})-\chi_{ij}(\mathbf{p})]^2=[\mathbf{q}\cdot\nabla\chi_{ij}(\mathbf{p})+\cdots]^2
\end{equation}
which is equivalent to multiply the cross section by the factor of
$1-\cos\theta$, since
\begin{equation}
1-\cos\theta=\frac{2q^2}{s}(1-\frac{\omega^2}{q^2}).
\end{equation}

Weldon investigated the decay rate in thermal field
theory\cite{Weldon}, pointing out that the statistical
distributions involving both direct and inverse reactions(detailed
balance) should be considered simultaneously, which naturally
demanded for introducing the transport cross section when
calculating the fermion damping.

Now we are aware of what has been lost in the straightforward
computation: that is $q^2$ in Eq.(\ref{kinetics factor q}), i.e.,
the statistical weights do not reflect the detailed balance, which
leads to an incorrect result. Therefore the right way to obtain
the effective fermion regulator(width) in transport process is to
employ the transport interaction rate and yields,
\begin{equation}
\gamma_e=\frac{1}{2}\Gamma_{trans}=\frac{1}{2}\int d\Gamma
(1-\cos\theta)
\end{equation}
and
\begin{equation}
\Gamma_{trans}(E)=\frac{e^2}{\pi v s}\int^\infty_0 dq\ q^3
\int^{vq}_{-vq}\frac{d\omega}{\omega}(1-\frac{\omega^2}{q^2})\left[\rho_l(\omega,q)+(1-\frac{\omega^2}{q^2})\rho_t(\omega,q)\right].
\end{equation}

Changing integration variables from $\omega$ to $x=\omega/q$, we
can evaluate the integral over $dq$ analytically and find out the
logarithmic dependence on the integral limits. This is again the
category which can be treated by employing the Braaten-Pisarski
method for the consistent leading order quantities, i.e. employing
a separating scale $\Lambda$ to divide the integral into a soft
contribution and a hard contribution, then adding the two
contributions which leads to the cancellation of  $\Lambda$.   One
would notice in the previous calculation on the photon damping
rate that the coefficients in front of the log-terms in the soft
and hard contributions are identical so as to guarantee the
separating scale to be cancelled. Thereby we can play a little
trick to obtain the leading-log result by just calculating the
soft contributions and replacing the separating scale with the
upper $T$, ignoring the constant factor added to the logarithm
term which is less important in the weak coupling limit.
Completing the integral over $dq$ we obtain\footnote{Since the
plasma is in a high temperature but small chemical potential
environment, i.e. $T\gg\mu$, the fermion lines in figure 4 should
be hard and the Pauli blocking has tiny effect on the final
result. Compared with the contribution from Bose-Einstein
distribution function of photon, the chemical potential in the
Fermi-Dirac distribution can be neglected. As to the density
effect in the full photon propagator, the small chemical potential
just modifies the constant in the logarithm but preserves the
order in $\alpha T$. Qualitatively one can adopt this result.}
\begin{equation}
\gamma_{e}=\frac{1}{2}\Gamma_{trans}(E)=\frac{3e^2T}{4\pi
s}m_\gamma^2\log\frac{T}{m_\gamma}.
\end{equation}
with $s=(P+K)^2=(2E_p)^2=4p^2$ in central mass system.

Combining ({\ref{fs2}}) with (\ref{dr}) and noticing that
$\gamma_e\sim e^4T \ll T$, one adopts a delta function as
approximation and finds:
\begin{equation}
G^e_R(Q)=\frac{2i\pi}{3}\int dK \frac{k^2}{\gamma_e}[\tilde
n_F(k_0)-\tilde n_F(k_0+q_0)][\delta(k_0-k)+\delta(k_0+k))].
\end{equation}
Integrating over $dk_0$ and then expanding the $\tilde n_F$ in
terms of small $q_0$, ignoring the terms without $q_0$(because Eq.
(\ref{v}) needs $q_0=0$ after differentiation on $q_0$), we obtain
\begin{equation}
G^e_R(Q)=\frac{2i\beta^2 q_0}{3\pi e^2 m^2_\gamma\ln e^{-1}}\int
dk k^6\left[\frac{e^{\beta(k+\mu)}}{(e^{\beta\mu}+e^{\beta k})^2}+
         \frac{e^{\beta(k+\mu)}}{(e^{\beta(k+\mu)}+1)^2}\right].\label{fermionv}
\end{equation}

Noticing $T\gg\mu$, one can expand the integrand of
(\ref{fermionv}) as Taylor series with respect to $\mu/T$ and
complete the final integration. The leading-log shear viscous
coefficient contributed by the fermions is:
\begin{equation}
\eta_e=361.4\frac{T^3}{e^4\ln\frac{1}{e}}(1+0.1765\frac{\mu^2}{T^2}).\label{fermionresult}
\end{equation}

On the contrary, if an interaction vertex connects the two
particles of different types, like Compton scattering or pair
production we have discussed in the previous section, no kinetics
factor $q^2$ will be isolated from the statistical weight, i.e.,
$1-\cos\theta$ will not appear in the photon width
expression\cite{Thoma,defu,Aarts3}. That is why we got the right
dependence of the coupling constant by just adopting the ordinary
photon damping rate as the infrared regulator in the photon
contribution. 

\section{Conclusions and Discussion}
Adding up the contributions from both the electron and photon
sectors, the shear viscous coefficient of QED plasma is obtained,
\begin{equation}
\eta_{QED}=\eta_e+\eta_{ph}=446.0\frac{T^3}{e^4\ln
e^{-1}}(1+0.1234\frac{\mu^2}{T^2}).
\end{equation}
One may find that the density-relevant part of viscosity coming
from the photons is negative but the one from electrons is
positive. They compete with each other and leave a positive
contribution of chemical potential, which demonstrates an
enhancement effect of density.

The above result is consistent, in the order accuracy of coupling
constant, but differing by a factor with those obtained in kinetic
theory at the finite density\cite{Liu} and the pure temperature
case\cite{Arnold1} when $\mu$ goes to zero. What we mean
consistency here is just at the qualitative level. More accurate
effect like ladder resummation, in which none damping rate should
be replaced by the transport one since it is self-consistently
including the main characters of the transport processes, should
be involved in the calculation. The one-loop approximation may
partly account for the discrepancy of the factors scaled by the
logarithms.

From the proceeding analysis we can conclude that the estimation
of the shear viscosity of QED plasma calculated at one-loop
skeleton diagram level by choosing proper fermion width in the
framework of the finite temperature field theory, is economical
and reliable. Other transport coefficients of the relativistic
plasma at finite temperature and density could be discussed
similarly. Using this method helps us to handle the pinch
singularity easily if the transport hard fermion and boson damping
rates are known. Nevertheless, we may point out that some further
improvements could be introduced to obtain more precise results
such as the Landau-Pomeranchuk-Migdal(LPM) effects due to the
interference of multiple scattering process\cite{Berges,Aurenche}
for the complete leading-order contribution.\\[1cm]

\centerline{\bf Acknowledgement} This work is partly supported by
the National Natural Science Foundation of China under project
Nos. 90303007, 10135030 and 10575043, the Ministry of Education of
China with Project No. CFKSTIP-704035. We thank M.H. Thoma for his
valuable comments.

 \end{document}